# Decentralized User-Centric Access Control using PubSub over Blockchain

Sayed Hadi Hashemi[1], Faraz Faghri[1], Roy H Campbell
*University of Illinois at Urbana-Champaign*

## Abstract

We present a mechanism that puts users in the center of control and empowers them to dictate the access to their collections of data. Revisiting the fundamental mechanisms in security for providing protection, our solution uses capabilities, access lists, and access rights following well-understood formal notions for reasoning about access. This contribution presents a practical, correct, auditable, transparent, distributed, and decentralized mechanism that is well-matched to the current emerging environments including Internet of Things, smart city, precision medicine, and autonomous cars. It is based on well-tested principles and practices used in distributed authorization, cryptocurrencies, and scalable computing.

## 1 Introduction

In not too long a fully deployed Internet of Things (IoT), smart city, precision medicine, and autonomous cars will become a reality, with it comes huge promises for the reduction of frictions and rise of efficiencies. This promise relies on analysis of massive amount of data, often times refereed to as big data analytics. However, gaining access to this massive amount of data, at a scale required by these environments, while preserving user's right is proving difficult. Data is stored in a variety of locations, managed by different service providers, to be analyzed by diverse entities, and subject to range of regulations concerning public records and individual privacy.

In order to make the dream of these emerging environments a reality, it is crucial to address the aforementioned challenges with more flexible and scalable data sharing models. Here, we present our solution, a user-centric data dissemination and distribution system to enable scalable data sharing and solving the following problems:

**1. User-centric.** data collected from the data sources potentially represents the activities of millions of users (individuals or organizations) and their possessions. Commonly, in these environments users themselves do not know what private data has been stored, who manages it, or where it resides. Arguably, this user factor should be informed and the ultimate deciding factor in control of their data.

**2. Privacy-preserving.** many users are willing to grant access to their data as long as their privacy is preserved. The goal here is to not allow data analyzers infer the identity of data owner through protocol mechanisms. The only way that an adversary can learn the identity of data owner would be through analysis of data content which is out of this work's scope.

**3. Endorsed.** naturally, trusted third parties should contribute in helping ascertain the trustworthiness of a party. But this role should be advisory, whereas now they are actively making these decisions on behalf of data owners.

**4. Accessible.** to take full advantage of data and produce actionable insight, data will need to be available for analysis to a diverse group of external sources such as researchers, social entrepreneurs, city departments, healthcare providers, and private sector. Also, any user such as data owners and data analyzers should be able to easily join the system.

**5. Scalable.** as the emerging environments scale to billions of data sources and millions of users, granting control to users becomes a very challenging problem. At this scale, conventional best practices are not practical, it becomes impossible to maintain an access control list on each single sensor, or near impossible to rely on centralized access control server, or agreed upon trusted parties based protocols. Networks of billions of devices and millions of users would require maintaining enormous accessible access control lists that if possible at all would be very expensive and challenging with their strict availability and consistency needs.

---
[1]Joint first authors

**6. De-centralized.** current environments, less and less frequently have centralized control, and are typically managed by a very diverse group of service providers. Because of these heterogeneous and disparate environments, it is often difficult to manage the issues of trust relationship management, data management, regulatory compliance and data sharing in a centralized control system. Server-client models such as Kerberos [5] can not be used in such environments due to lack of central point of trust.

**7. Distributed.** data resides on the clouds and at the edge on billions of data sources. As the emerging environments grow and data generating devices deployed at ever increasing rates, we find that data remain siloed and inaccessible. Mainly due to the fact that the generated data is owned by millions of users and managed by a wide range of organizations.

**8. Asynchronous.** lack of connectivity is a traditional challenge in many environments. Devices might not be nearly as well connected. A network of sensors implemented in a house that is not necessarily exposed to the whole world. This requires systems to function without the need of persistent connectivity.

**9. Compliant.** many countries proscribe specific requirements and liabilities upon organizations that hold or manage data. These entities are often subject to the regulatory requirements, e.g. HIPAA, Directive 2011/24/EU, and other similar laws for entities covered by healthcare privacy regulations.

**10. Transparent.** in the absence of a centralized trusted party, transparency is a key principle to keep parties honest. Enhancing transparency ensures that legislation are adhered to in the emerging environments.

**11. Auditable.** users should be able to share their data with any chosen data requester, as well as track and monitor the requesters access. Audit is a critical feature to track who, when, and where has access a data.

**12. Correct.** showing security correctness is a major requirement for practical applications in order to prevent protocol and implementation flaws.

**13. Usable.** we need a system that can securely provide the above functionalities for locating and sharing data between data sources, and/or entities. Distribution and privacy concerns make implementation of very basic Big Data Analytics primitives nearly impossible. For instance, in the current setting, 'dataset lookup', the simple task of finding a data set is a major quest.

**Our solution** provides a mechanism for the user to control at scale. Two differentiating contributions in our solution are: first, separation of the data store from the data management, and second, architecting both components scalable, decentralized and distributed. These two major contributions provide substantial benefit to resolving aforementioned challenges; it grants the possibility of keeping data at its origin, overcomes the single point of trust and failure, and adapts to the growth of users and data sources.[1]

**This paper** describes and evaluates a practical and correct implementation of this solution in a system called Dusc. The paper contains an improved version of the system, a full discussion of an implementation, a proof of the system's correctness, and an evaluation of its practicality by performance measurement, storage costs, query times, and Blockchain consistency times. This version provides substantial improvements including support for local parallelization.

**Dusc** has three main components which utilizes best practices in computer security and distributed systems: data management protocol, messaging service, and data store system. First, a data management protocol in which different roles can interact with each other in a secure and private manner based on capability-based access control. Second, a decentralized and distributed data store system that allows each role able to communicate with each other without the need for a trusted third party based on Blockchain. Although our solution has a different purpose, recent cryptocurrencies such as Bitcoin provide an interesting and similar successful implementations of decentralized and secure circulation of currency. Third, a scalable messaging service based on publish-subscribe model designed to provide flexible and reliable communication between many senders and receivers without the need of persistent connectivity. Our solution is fully decentralized and distributed in order to accommodate the massive scale of the IoT, smart city, precision medicine, and autonomous cars paradigms. Because the system is user-centric, no trust of any participant is implied and automatic. While still maintaining the ability to scale, this provides full control of access control and propagation, system security, and user privacy.

---

[1]Preliminary version of this idea in the context of Internet of Things has recently been proposed by the authors. The manuscript will appear soon in an IoT conference. Due to the "Anonymous submission" policy, paper is not referenced. Contribution of the current paper compared to the IoT one is highlighted in the next paragraph.

The remainder of the paper is organized as follows. In Section 2, we give an overview of our solution. Sections 3, 4, and 5 will describe the three components of our solution, data management protocol, data store system, and messaging service respectively. We discus the implementation and evaluation of our proposed solution in Section 6. Section 7 looks back at how Duschas addressed the aforementioned challenges.

## 2 System Model and Design

Further, we explain the roles, primitives offered to these roles, along with the scalability and threat assumptions. For the rest of this paper we use Abadi et. al notations [2],[3] to formally describe the system and protocols. Notations are explained in Table 1.

### 2.1 Roles

Four main roles are introduced in our model:

- **Data Owner:** an individual or organization who is in possession of data. This role does not necessarily generate or store the data. In our model, data owner grants the access to the data.

- **Data Source:** represents a computer system, individual, or organization, who manages and stores data objects, be it at rest or in motion. Examples would include cloud providers, managers of Electronic Health Record (EHR) systems, application gateways, and archival systems. A sensor can act as data source if it has enough performance, connectivity, and storage, otherwise its data is stored elsewhere.

- **Data Requester:** an individual or organization that requests access to other data owners' data, available within the network. Examples are researcher, company, data aggregator, or another device.

- **Endorser:** a third party individual or organization that validates a request. This may be a trusted authority, organization, or known individual. Endorser either provides supplementary information regarding credibility, or validates authenticity of the role's identity. Examples of endorser are but not limited to: FDA, NIH, DMV, municipal, Internal Review Board (IRB), and external review by additional research institute (Hospital A asks hospital B for review).

### 2.2 Primitives

Following primitives are considered for the above roles in our model:

- **Data Discovery:** users have many data generating devices and more likely no substantial data storage. They should know what data they own or have the rights to. They should have one view that allows them to see a portfolio of all of the data they own.

- **Data Request:** the subject should be able to search for a collection of data that meets their conditions. The request should be easily converted to a data access authorization.

- **Audit:** users should be able to share their data with any chosen data requester, as well as track and monitor the requester's access.

### 2.3 Assumptions

Following scalability and threat assumptions are considered in our model:

A1. We assume that data sources are aware of their users' identity (public key), i.e. data source $S$ knows its user's public key $K_X$, and can securely authenticate signed messages by that user:

$$\{message\}_{X'} \implies X \text{ says } message$$

A2. We assume that during user registration, users are provided with a mechanism to authenticate the identity of the data source. This mechanism can be verification from a certificate authority or a token:

$$\{K_C\}_X \implies C$$

A3. We assume that the endorsers' identity can be relied upon as valid via currently available technologies such as PKI, i.e. if a subject trusts a

Table 1: Notations used in this paper adopted from Abadi et. al [2],[3]

| Notation | Description |
|---|---|
| $\{M\}_X$ | Encrypted message *M* using *X*'s public key. |
| $\{M\}_{X'}$ | Signed message *M* using *X*'s private key. |
| $X$ | Identity of *X*. |
| $K_X$ | *X*'s public key. |
| $O_i$ | Data Owner *i*. |
| $S_i$ | Data Source *i*. |
| $R_i$ | Data Requester *i*. |
| $E_i$ | Endowser *i*. |
| DOT | Data Object Ticket. |
| DAP | Data Access Path. |
| RT | Request Ticket. |
| DAT | Data Access Ticket (Capability) |
| *X* says *Y* | *X* makes the statement *Y*. |
| *X* for *Y* | *Y* on behalf of *X*. |
| *X* controls *Y* | *X* is trusted on *Y*: if *X* says *Y* then *Y*. This is the meaning of *X* appearing in the ACL for *Y*. [2] |

certificate authority such as *C* which signs an endorser identity *E*, then:

$$K_C \implies C$$
$$K_C \text{ says } (K_E \implies E) \to K_E \implies E$$

- A4. We address the issue of traffic analysis in this paper; however we make a best effort attempt rather than attempting to completely obfuscate users' access patterns.

- A5. It is assumed that users may trust their local computing environment. We understand that this assumption is difficult to guarantee and the execution of sensitive programs in untrusted environments will continue to be a risk.

- A6. We assume that attackers have a specific set of abilities. We assume that attackers can view system's data, are able to present modified data to participants, and may impersonate roles. We also assume that attackers do not have access to private keys, and cannot gain local administrator access to the systems.

- A7. We assume the security of the Blockchain system, and cryptographic integrity of the utilized encryption protocols.

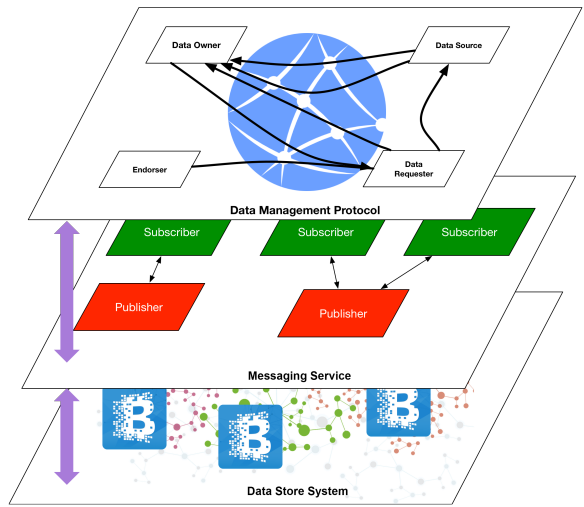

Figure 1: Our solution consists of three main components: *i.* data management protocol, *ii.* data store system, and *iii.* messaging service. Data management protocol is a secure capability-based system. The messaging service is based on publish-subscribe architecture. Data store system is based on the Blockchain.

- A8. We assume that 'things' either have the capability to act as a data source or are able to transmit their data to a data source securely.

- A9. We assume that each data object has at least one data owner.

- A10. We do not address DoS attacks in the current system. In the future works, this attack can be addressed by the best efforts and practices.

## 2.4 Our solution

Reviewed the roles, primitives, and assumptions in our model, we introduce our solution for empowering the users by controlling the access to their data at scale. Our system consists of three main components:

1. Data management protocol
2. Data store system
3. Messaging service

An overview of our system with the three components is illustrated in Figure 1. Data management protocol provides a framework for interaction among different roles with an access control mechanism. The protocol is a secure capability-based system. Data store system provides a persistent, distributed, and decentralized storage for the access control component; it is based on Blockchain and provides full transparency of transactions very similar to Bitcoin. The messaging service plays an important role for the scalability of our solution; it is based on publish-subscribe architecture and designed to provide scalable, flexible, and reliable communication between many senders and receivers without the need of persistent connectivity.

Our solution is user-oriented; users are the bridge between different networks of data generating devices, and they are ultimately authorizing the access to their data. Information regarding the user's possessions is sent to the user securely and privately. This data is accessible whenever the user demands to view it. Data owners are aware of the existence of data by direct knowledge or control over the sensor/actuator producing the data. Additionally, they may learn about the existence of data via notifications generated by the data sources and transmitted directly to the data owner, mainly to communicate the management of data by the data source.

Data requests are broadcasted to all the data owners within the system. Data owners are not required to review all the requests. In case a data owner is not interested in sharing any data, they can easily filter out all the requests, or they can selectively filter out requests based on their interest. Also, user's trusted endorsers may verify a data request. This helps protecting users against malicious or unethical data requests, as well as assisting them with risk assessment of received data requests.

Data requests, upon arrival by the user, will be checked against the user's portfolio in the client. If the conditions are met, and data owner is willing to share the data, a capability ticket will be issued to the data requester which provides necessary authorization for accessing the requested data.

Whenever two roles want to interact, one sends a message to the messaging service component. Then, the elements of messaging service will send the message to the data store system which is based on Blockchain. After successful storage of the message in the data store system, it will be fetched by the recipient through the messaging service.

In the next sections, we explain these three components in more detail.

## 3 Data Management Protocol

Data management protocol provides a framework for different roles to interact in a secure and private manner. The protocol is a decentralized capability-based access control. The communication between different roles is performed by message passing. Each message contains a sender, a receiver, and a payload. Messages are delivered through the underlying "Messaging Service" which is described later in Section 5.

In this section, first, we compare our data management protocol to capability-based access control systems. Later, we introduce data structures used in the message payloads along with the messages transmitted in the protocol. Throughout this section, we use notions described in Table 1.

### 3.1 Capability-Based Access Control

Our system is implementing a distributed and decentralized capability based access control system. Capability is a token that gives a data requester permission to access a data owner's data object on a data source. Using capability based access control, our system avoids the necessity of having a centralized trusted party to confer trust. Instead, the data owners are responsible for issuing the capability to other users.

In our implementation, *capabilities* are issued by the data owners ($O$). Each one indicates who is the issuer, to which data requester ($R$) this capability is issued, and the object to grant access to. Additionally, each capability contains the access rights in the form of data queries, it also contains information about where and how the data can be accessed. Capabilities are not transferable or usable by any subject other than the intended data requester. As the result, the data owner will remain the sole controller of the data:

$$(R \text{ for } O) \text{ says } d$$

Our capability is implemented as a data structure that contains:

- **Access rights:** every capability issued contains a query that adds additional restrictions on the data access.

- **Identities:** are used to uniquely describe capability issuer and issue. Due to the decentralized nature of the system, it is not an easy task to identify every subject. Therefore, public keys are used to identify subjects in the capability. Based on implementations such as RSA, or PKI, it is assumed that generated key pairs are always unique. There is a many-to-one relationship between capabilities and a public key. Multiple entities can share a public/private key pair in order to implement a joint ownership.

$$K_X \Rightarrow X$$

In our implementation of capability, following operations are contained:

- **Create:** right to create capabilities is restricted to the data owners (*O*). As soon as the data object (*d*) is generated and stored on a data source (*S*), this right is delegated to data owners in order to grant access to potential data requesters (*R*). This creation of right is not transferable.

$$\forall R, O \text{ controls } [(R \text{ for } O) \text{ says } d]$$

- **Delegate:** deleting a capability is not possible for the sake of audition and non-repudiation. However, a data object referred by the capability can be revoked by moving (or removing) the data on the data source.

**Audit** is a critical feature to track who, when, and where has access a data. In the absence of a centralized trusted party, our implementation provides the audit feature through the capability call back key. Data sources will inform the data owners when capabilities are used to access data objects.

Also, the capability is stored by the data requester. Upon arrival, capability can be verified without an external *access control list*. As a result, having more capabilities or data objects may not require extra storage and computing power on the data storage, nor the data owner.

## 3.2 Protocol Data Structures

Five data structures are used in the payload of messages. *Data Objects* are data at rest or data in motion stored in the data sources. Even though the data objects are not directly used and transmitted in the message, the ultimate goal is to manage their access. Data objects could be files or records within a file repository, database, cloud provider, or any other data storage system. In practice, there are no limiting factors that would prevent our system from supporting other documents such as paper documents, film, recordings or another non-digitized media. Data Objects in motion could be a variety of live data sources, including but not limited to: sensors, video streams, audio streams, health trackers, location data, etc. Lastly, a Data Object may represent an actuator or device that the data requester may interact with. An example may be a Physical Access Control Systems (door locks, and alarm systems), a PTZ camera, a traffic light, a shared vehicle ignition system, etc.

Five data structures used in the payload of messages transmitted in the Data Management Protocol are:

1. **Data Object Ticket (DOT):** the right to issue a capability to access a data object. This ticket is issued by the data source and contains the unique id of specified data object, the identity of the owner ($K_O$), the identity of the data source ($K_S$), data access path (*DAP*), and metadata describing attributes of data object, signed by the data source:

    $$\text{DOT} = \{\text{Data ID}, K_O, \text{metadata}, \text{DAP}\}_{K'_S}, K_S$$

2. **Data Access Path (DAP):** this message represents a mechanism for an authorized role to gain access to the data object. Some examples of this DAP can be:

    - URL/URI (FTP/SFTP/SCP/HTTP/HTTPS)
    - Record Locator
    - Contact Information
    - Instructions
    - Physical Location

    Note: In our implementation, tickets can be utilized as the authentication mechanism for TLS/SFTP/SSH as they are generated from the TLS standard.

3. **Request Tickets (RT):** these are messages broadcasted by the data requester to all the subscribers. A data request ticket typically consists of a data query, participation conditions, duration of access, and some relevant metadata:

    $$\text{RT} = \{\text{Request ID}, K_O, \text{Query}, \text{Conditions}, \text{Duration},\\ \text{Metadata}, K_R\}_{K'_R}$$

    A data query indicates to the data owner what data the data requester intends to collect from data owners. The query format is determined by data sources. Participation conditions are constraints on data owner in order to determine if they are qualified to participate in the data request. For example in a medical data collection this condition can be on:

- Nationality of Participant
- Demographic information about participant, age, gender, or residency
- Relationships with certain data source providers, i.e. only data stored with Google or Apple

4. **Endorsements:** provide a method for an endorser or endorsers to provide additional third party information to the data request. This information will help the data owners to decide whether they are willing to share data for a request. For example, endorsement can be:

    - Metadata regarding the data privacy policy of the data requester
    - Results of a third party audit or attestation regarding the information security controls in place at the organization
    - Results of an Internal Review Board (IRB) decision
    - Authorization by a government entity to conduct a trial or research (FAA, NIH, HHS)
    - Indication that data will be shared in a joint collaboration
    - Verification of the data requester authenticity by a service provider or other trusted third party (Think CA/Trusted Roots in PKI)

    Endorsement is done by chain of signatures. For example the following request ticket is endorsed by two endorser $E1$ and $E2$:

    $$\{\{\text{RT}, K_R, Feedback_{E1}\}_{K'_{E1}}, E1, Feedback_{E2}\}_{K'_{E2}}$$

5. **Data Access Ticket (DAT):** is a capability that gives authorization to access a data object. The ticket is issued by the data owner to grant the specified data requester access to the data source and contains the identity of the entity who will access the data. But it is not sent directly to the data source. Additionally it contains an identity to receive the acknowledgment when the ticket is being used.

    $$DAT = \{K_O, DOT, \{Data\ ID, Query,$$
    $$K_O, K_O^3, K_R\}_{K'_O}\}_{K_S}$$

In order to protect the real identity of data owners, a data owner may choose to use multiple identities during the ticket exchange process:

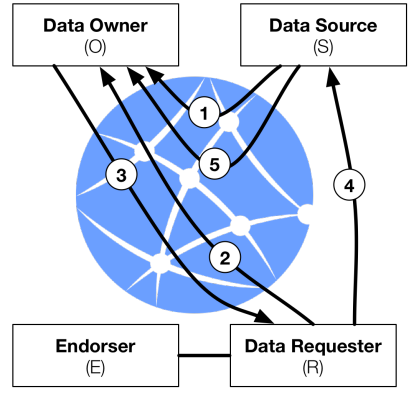

Figure 2: Data Management Protocol with five class of messages transmitted among four roles. Numbers on connections correspond to the message types.

- $K_O$ is the identity that is shared with the data source
- $K_O^2$ is the identity to be used to communicate with the data requester. This identity is not known by the data source
- $K_O^3$ is the identity to receive the acknowledgment of ticket being used. This identity is not known by the data requester

## 3.3 Authorization Messages in the Protocol

In this section, we describe the messages transmitted in the protocol. These five class of messages are illustrated in Figure 2 and a summary is provided in Figure 3. Messages contained in the protocol are:

- **Message 1: Data Source Ticket Generation** The goal of this message is to let the data owner ($O$) know about the existence of a new data object which belongs to the recipient. Furthermore, the included data object ticket allows for the owner to delegate discretionary access to the data object.

    The other included token $\{K_S\}_{K'_O}$ (from $A1$ assumption) authenticates the data source to the data owner. Later, *DOT* can be used to authenticate the recipient.

- **Message 2: Data Request** The data requester broadcasts a data request to the system subscribers. This data request may contain endorsements from one or more endorsers which may provide supplementary information regarding the data request.

$$(M1)\ S \to O : \{DOT\{K_S\}_{K'_O}\}_{K_O}$$
$$DOT = \{Data\ ID, K_O, metadata, DAP\}_{K'_S}$$

$$(M2)\ R \to * : \{[\{Hash\{RT\}, Feedback\}_{K'_{E1}}]^+,$$
$$RT, K_{E1}, K_R\}_{K'_R}, K_R$$
$$RT = \{Request\ ID, Query, Conditions,$$
$$Metadata, Duration, K_R\}_{K'_R}$$

$$(M3)\ O\ via\ K_O^2 \to R : [\{Data\ ID, DAP, DAT, K_S\}_{K_R}]^+$$
$$\{Request\ ID, K_O^2, \{Checksum\}_{K_O^{2'}}\}_{K_O^{2'}},$$
$$DAT = \{K_O, DOT, \{Data\ ID,\ Request\ ID,$$
$$Query, K_O, K_O^3, K_R\}_{K'_O}\}_{K_S}$$

$$(M4)\ R \to S : [DAT]^+,$$
$$\{\{Hash\{[DAT]^+\}\}_{K'_R} K_R\}_{K_S}$$

$$(M5)\ S \to O : \{K_S, DOT,$$
$$\{Query, K_O, K_O^3, K_R\}_{K'_O}\}_{K_O^3}$$

Figure 3: Five access control messages in the Data Management Protocol.

The data owner's client will process the data requests against its library of data object tickets in order to check if participation condition is met, and identify which tickets may meet the specifications of the data query.

Endorsements as well as supplementary information can also help the data owner to decide on whether to participate in the data request. Data owner can verify originate of endorsement by checking endorser's identities (from $A3$ assumption) as well as confirming the $K_R$ in $RT$ is matched with the sender of request.

- **Message 3: Ticket Exchange** When the data owner consents to grant access to the data object(s), a ticket exchange message is sent to the data requester. This ticket contains one or more data access tickets per data source, each one containing part of requested data. Each data access ticket includes identity $K_S$ of the data source which is storing the data object, as well as the address that the data can be accessed by (*DAP*).

- **Message 4: Data Access** As as a result of the message 3, the data requester has now received the data access path as well as any other relevant information needed to access the data. Depending on the application, the data requester may contact the data source(s) directly or indirectly through the system.

  Data source can verify the access by testing:

  1. *DOT* is signed by the data source *S* and includes $K_O$. This will verify *O* right to grant a capability on *Data ID*:
     *DOT* $\Rightarrow$ *O controls Data ID*

  2. *DAT* includes signed *Data ID*, *Query*, and $K_R$ with $K_O$ which matched with requester key. This verifies *O* grant a capability on *Data ID* using *Query* for *R*:

     $\{K_R, Data\ ID, Query\}_{K'_O} \Rightarrow$
     $O\ says\ (K_R \wedge\ Data\ ID\ \wedge\ Query)$
     $(R\ for\ O)\ says\ DataID \wedge Query$

  3. Optionally, the data source can check *Request ID* against a data request black list.

- **Message 5: Access Announcement** In step 5, it is expected that Data Sources will announce to the system that access has been made utilizing the DAP(s) and credentials provided in Step 3 and accessed as part of Step 4. This allows the data owner to monitor accesses to their data and inform other parties about successful transmission of transaction.

## 4 Data Store System

In this section we explain the design of the access control data store system. This storage system supports the distributed access control and is separated from the source data. Separation of source data grants the possibility of keeping source data at its origin. The access control data store component is a persistent, scalable, decentralized, and distributed storage system based on Blockchain. We model the Blockchain as a form of persistent data storage with update notification support. It overcomes the single point of trust and failure, and provides full transparency of transactions very similar to Bitcoin. It has to be noted that the Bitcoin's Blockchain is just an example of an appropriate protocol, possibly, there exists other such appropriate protocols. Before explaining the system, it is necessary to review core components of the system which are adopted from the Bitcoin's Blockchain. Then we show how Blockchain can be used as data storage:

## 4.1 Blockchain

Bitcoin, the system first introduced by Nakamoto [4], is the first truly decentralized global currency system. Like any other currency, its main purpose is to facilitate the exchange of goods and services by offering a commonly accepted value. Unlike traditional currencies however, Bitcoin does not rely on a centralized authority to issue, control the supply, distribution and verification of the validity of transactions. Bitcoin enables a network of computers to maintain a collective bookkeeping via the internet. This bookkeeping is neither closed nor in control of one party. Rather, it is public and available in one digital ledger called Blockchain which is fully distributed across the network and relies on a network of volunteers that collectively implement a replicated ledger and verify transactions. Traditionally, Blockchain has been discussed in the context of Bitcoin, however Blockchain goes beyond the scope of consensus currency, introduces many new and innovative methods for propagating information in the network, public transaction history, multi granularity and many others.

Blockchain uses a multi-hop broadcast to propagate transactions and blocks through the network to update the ledger replicas. In the Blockchain, all transactions are logged including information on the date, time, participants and amount of every single transaction. Each node in the network owns a full copy of the Blockchain and on the basis of cryptographic principles, the transactions are verified by the so-called Bitcoin Miners, who maintain the ledger. The systematic eventual consistency principles also ensure that these nodes automatically and continuously agree about the current state of the ledger and every transaction in it. If anyone attempts to corrupt a transaction, the nodes will not arrive at a consensus and hence will refuse to incorporate the transaction in the Blockchain. So every transaction is public and thousands of nodes unanimously agree that a transaction has occurred on particular date and time. In Blockchain, trust comes from the fact that everyone has access to a shared single source of truth. The decentralized property is the fundamental difference from previous systems which relied on a centralized block issuer and required users to trust the original issuer, which was still used to eventually clear transactions.

In our solution, we use the Blockchain to store the access control data in a decentralized manner. Prior to describing the decentralized access control, we need to discuss the Blockchain as data storage.

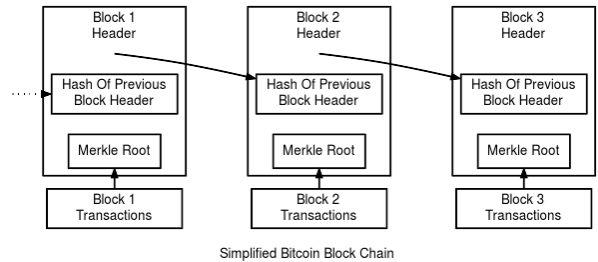

Figure 4: Overview of the Blockchain. Every block contains a hash of the previous block. New transactions are constantly added to the end of the chain, source[4].

## 4.2 Data Model in Blockchain

A block chain is a transaction database shared by all nodes participating in a system. As pointed out earlier, Blockchain can be used as data storage for many different applications; it provides various storage functionalities, among those three primitives are essential to our system: *i.* retrieve, *ii.* update, and *iii.* add. Further, we explain how these primates are supported in the Blockchain and form the data flow, but first we review some definitions in the Blockchain adapted from Nakamoto paper [4] and the Bitcoin Developer Guide [1], illustrated in Figure 4:

- **Transaction:** a transaction is a transfer of value (e.g. Bitcoin, information) that is broadcast to the network and collected into blocks. A transaction references previous transaction outputs as new transaction inputs and dedicates all input Bitcoin values to new outputs. It is possible to browse and view every transaction ever collected into a block.

- **Block:** transaction data is permanently recorded in files called blocks. Blocks are organized into a linear sequence over time (also known as the block chain). As Figure 4 illustrates, every block contains a hash of the previous block. New transactions are constantly being processes by miners into new blocks which are added to the end of the chain and can never be changed or removed once accepted by the network.

- **Mining:** is a distributed consensus system that is used to confirm transactions and add transaction records to the public ledger of past transactions (Blockchain). It enforces a chronological order in the Blockchain and allows different computers to agree on the state of the system.

- **Miner:** is an individual or an organization performing the mining. Miners dedicate considerable computation power for maintaining the Blockchain. In Bitcoin miners are incentivized by a reward i.e. Bitcoins. In our system miners are researchers and organizations requesting and analyzing the data; it is in their benefit to mine the Blockchain.

- **Blockchain:** is a transaction database shared by all nodes participating in a system. A full copy of a Blockchain contains every transaction in order, dating back to the very first one. The entire Blockchain can be downloaded and openly reviewed by anyone.

- **Genesis block:** is the first block of a Blockchain. The genesis block is hardcoded into the software and is a special case in that it does not reference a previous block.

Our system's data flow is based on three essential primitives enabled by the Blockchain:

1. **Retrieve:** per design, the entire Blockchain can be retrieved and downloaded by anyone. With this information, one can find out how much value (e.g. Bitcoin, information) belonged to each entity at any point in history.

2. **Update:** transactions and mining results are broadcasted in the network, every new block is ordered and linked to the previous block which makes it impossible for nodes to miss any added information.

3. **Add:** adding data is the same process as transferring data:

   (a) When a node in a Blockchain wants to transfer data, the node broadcasts the request, the request is received by all the nodes on the Blockchain network.

   (b) After receiving the request, nodes which are miners will add this most recent transaction request into a block. Then they run the new block and the previous block into a set of hash function based calculations.

   (c) All the miners start racing on the complicated cryptographic puzzle. When the first miner solved the block, it adds the block to the end of Blockchain and will broadcast it to its peers.

   (d) After the broadcast, peers will check the transaction and will start using the new version of Blockchain.

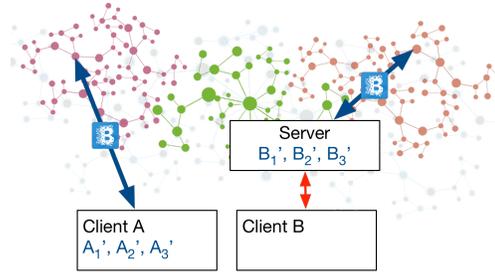

Figure 5: Client Access Models to the BlockChain. Left: Direct Access, Right: Server Client. Blue lines indicate Blockchain access, red line indicates API call.

In the next section, we discuss different mechanisms to allow our clients to connect to the data store system. We use the above primitives and implement a publish-subscribe messaging servie which delivers transactions to their intended recipients.

## 5 Messaging Service

In all the emerging big data analytic domains, interaction with the data flow is a major issue and it is crucial to adopt the most scalable solution. In general, there are three different client data access models available for users to access and communicate with the data store system: *i.* direct access, *ii.* server-client model, and *iii.* publish-subscribe model.

In this section, we review these three options. We explain how publish-subscribe architecture provides scalable, flexible, and reliable communication between many senders and receivers without the need of persistent connectivity. Furthermore, we explain how we adapt the publish-subscribe model into our solution. In our system, whenever two roles want to interact, one sends a message to the messaging service component. Then, the elements of messaging service will send the message to the data store system which is based on Blockchain. After successful storage of the message in the Blockchain (data store system), it will be fetched by the recipient through the messaging service.

### 5.1 Data Access Model

In order to address a variety of use cases we present three different client data access models for data owners and other participants to interact with the Data Flow.

- **Direct Access:** Blockchain openness allows anyone to download the whole distributed database up to the present and subsequent updates as deltas (blocks). (Figure 5-Left) This method is very straightforward and requires minimal implementation. It provides full validation of the Blockchain, as well as maintaining the highest level of safety and privacy. However, this model is not feasible to implement in all environments. This model requires a powerful, always-on computer to handle a large amount of data and to query this data. Devices such as mobile phones, embedded systems or other less powerful systems typically lack the resources to interact directly with the data flow. In reality, each entity whether it is Data Owner, Data Source, or Data Requester, is interested only in a small fraction of data. Excess amount of redundant processing, as each client will need to process and download the entire Blockchain, results in a high amount of waste.

- **Server-Client:** In environments where there is a high degree of trust between two entities a client-server model avoids the redundant processing associated with the Direct Access Model. In a client-server model the server handles all of the client's interactions with the data flow (Figure 5-Right). In these environments the client's public and private keys are transmitted to the server which acts on the client's behalf. Typically the client's interaction with the data flow will be completely abstracted by an application that interprets communication received from the server. Some strengths of this model include: it allows for password resets or other ways for the server to validate the client's identity, in case the client loses its keys. This model introduces risks into the system. Since client's keys are stored on the server, in a server-client environment, if a server is compromised, a client's keys might also get stolen. Since these servers typically store large number of keys, they are often the target of attackers.

- **Publish-Subscribe (Pub-Sub):** This model represents a mixture of the previous two systems. In a Pub-Sub system the Publisher or Server monitors the data flow on behalf of the Subscriber (Figure 6). This substantially minimizes the processing load placed on the data owner. Clients subscribe to a set of queries. The publisher then filters out incoming traffic based on these queries, and only communicates matching queries to the subscriber. In these systems, the publisher does not have access to the data. It communicates the encrypted information, via a subscription, to the subscriber who is then able to decrypt the information with their private key. The traffic will be decrypted and processed on the client side. The only query required in the protocol is matching the destination to an identity. This query can be implemented very efficiently using Bloom filter and provided by the pub-sub model.

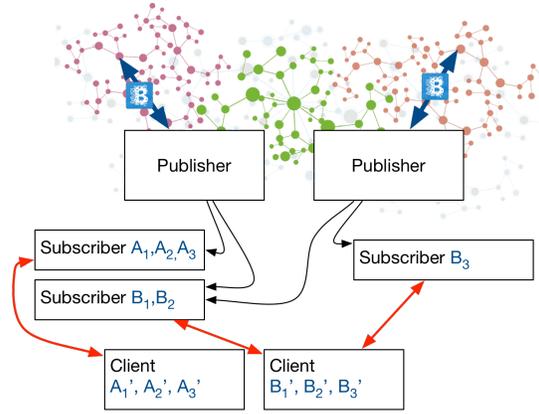

Figure 6: Publish-Subscribe client access model to the Blockchain. Blue lines indicate the Blockchain access, red lines indicate API call, and black lines indicate publish-subscribe data transfer.

Next, we explain how we adopt the publish-subscribe model into our solution.

## 5.2 Publish-Subscribe Model

In this section, we explain how users' clients utilize the subscribers to receive specific updates from the Blockchain. As explained in the previous section, a Blockchain is a collection of blocks. Each block contains a set of transactions such as ticket request and data access, between different roles in the system. By using a publish-subscribe model, publishers join the Blockchain, collect and filter appropriate transactions, and provide subscribers with their requested transactions. This process includes the following mechanisms:

1. **Join:** after joining the Blockchain, each publisher receives updates and will be able to access previous blocks in the Blockchain. Then, subscribers subscribe to one or more publishers, request a dedicated cache space to be initiated on the publisher, and provide them with a set of identities (public keys). The set of identities determines which transactions the subscriber is interested to receive. The cache space is intended to store data so data can be served faster when the subscriber requests it. It should be noted that publishers are located on machines that have access to the Blockchain updates, and also have stored copy of the Blockchain for fast local access. Subscribers are located on the client's machine.

2. **Receive updates:** whenever a new subscriber is added, it is typically specified from which past block it would like to start receiving new blocks. Because of this, publishers can ensure that they are sending updates beginning with the correct block. Ideally all subscribers will receive updates from the same block, i.e. the last added block. Hence, publishers use one reader to retrieve updates, and then relay them to all interested subscribers. In the case of error conditions, the publisher needs to restart sending updates from the last acknowledged location. For this, the publisher may need to read an older update. If many subscribers are recovering simultaneously, a naive implementation, a publisher would read from many positions in blocks simultaneously, which might cause high publisher I/O load. Another approach is to have one recovery thread from the oldest position working towards the end.

3. **Filter:** after receiving the updates from the Blockchain, publisher breaks down the blocks into original transactions. Then, transactions are filtered on the publisher side; the subscribers inform the publishers of what filters they need. The Publisher only delivers updates that meet the specifications of the supplied filters. While the evaluation of filters does place some additional processing overhead on the publisher, it helps conserve both memory and network bandwidth. This is especially the case when there are many subscribers that require only a small subset of the data.

   Subscribers request filtering for a set of identities that may be matched with either the sender or recipient of a transaction. For example, a subscriber may request that a publisher filters messages only set as broadcast to all members. Using this structure we can easily implement a Bloom filter, which works efficiently under these conditions. If a transaction is matched, it will be kept and stored in the cache corresponding to the subscriber.

4. **Deliver:** reliability is an important requirement. For example, one missed update could lead to permanent corruption in a user's portfolio. Subscribers receive the update stream of blocks. Publishers periodically track the delivery of updates by having subscribers acknowledging the delivery of updates. It is assumed that Subscribers are stateless, i.e it is not required for them to keep track of the state of deliveries. When a subscriber asks for receiving its updates, publisher sends all the transactions in the cache corresponding to the subscriber and asks for the subscriber's acknowledgment. Upon receiving the acknowledgment, publisher clears the cache. Otherwise, publisher keeps trying upon receiving the acknowledgment.

It should be noted that one client may deploy many subscribers, and one subscriber could join many unique publishers. Each publisher joins only one Blockchain.

# 6 Implementation and Evaluation

To estimate the performance of Dusc, we conduct a number of experiments to collect real performance characteristics of our prototype implementation. Our experiments are performed in the following settings:

- **Server**. we use a 2.2 GHz Intel(R) Xeon(R) CPU E5-2660 Machine with 128GB of Ram, with Ubuntu 14.04 x64 OS. Codes on this platform are developed using Python 2.7 and PyCrypto library for cryptographic algorithms.

- **Client**. we use an iPhone 6 with iOS 9.2.1. Codes on this platform are developed using Swift 2 and XCode 7. Apple Security Framework and Heimdall library is used for cryptographic algorithms.

To minimize the error and increase the confidence level, each experiment is repeated 100 times and the median measured value is reported. Further, we describe the performance of three components of Dusc:

### 6.0.1 Data Management Protocol

**Authorization costs**. as mentioned earlier, due to scalability challenges, it is not feasible for data sources to maintain an ACL. Therefore, data sources compute the authorization upon receipt of data requests rather than performing a lookup. To evaluate the overhead of this operation, we execute a batch of authorization checks on data requests with variable number of capabilities. Each request could contain multiple capabilities, the time spent to process one capability is $3.53e-2\ s$ constant. Figure 7 shows the measured time for authorization check of a data request with different number of capabilities on the server with single core resource allocation. This metric performs linearly in respect to the number of capabilities, but constant in respect to the number of owners and data objects. For a request with 1M capability, authorization takes $2.6\ s$.

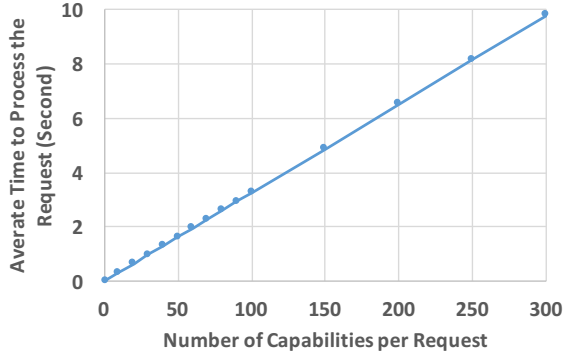

Figure 7: Authorization costs in the data management protocol.

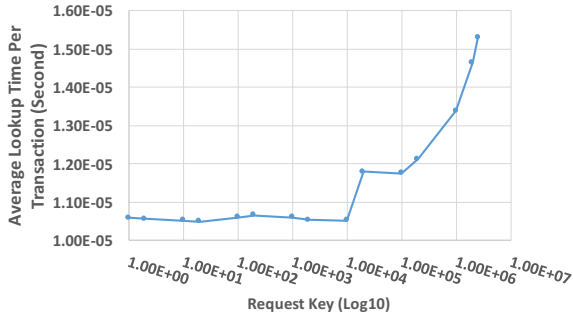

Figure 8: Bloom filter performance in messaging service Pub-sub.

**Create and verify messages.** data management protocol contains five different class of messages. For each message communicated, the sender creates a message and the recipient verifies the integrity of the message. We have evaluated these overheads for three different roles: data source, data owner, and data requester. We have also considered two scenarios where each role is executed on either the server or client side. As shown in Table 2, the overhead in all the above scenarios is negligible.

#### 6.0.2 Messaging Service

**Pub-Sub performance.** persistency of messages and scalability of messaging service in Duscrelies on the performance of the implemented Pub-Sub. For our Pub-Sub implementation, we use Bloom filters implementation of [6] to achieve near linear filtering performance. Figure 8 shows time spent to filter transactions in respect to the number of requested keys. For each transaction in Dusc, there exists two keys corresponding to sender and receiver. We use RSA generated 1024-bit public key. On average, it takes $1.12e-5\ s$ to filter each transaction, when 1M keys are presented under 0.1% Bloom filter error rate. This time is almost negligible in respect to increased number of keys.

## 7 Discussion

In this section we look back at the problems facing scalable data access and how Duscaddresses the aforementioned challenges.

The following challenges have been addressed through the data management protocol:

**1. User-centric.** users are in full control of granting data access to their collective data. They can always query the system and discover data that belongs to them.

**2. Privacy-preserving.** users preserve their privacy by consciously choosing who can access to what data.

**3. Endorsed.** a well-defined role with well-defined responsibilities.

**12. Correct.** we have shown the correctness in Appendix I.

The following challenges have been addressed through the messaging service and data store system:

**4. Accessible.** every individual or organization can easily join the Blockchain and Pub-Sub system.

**5. Scalable.** scalability performance have both components have been discussed in Section 6.

**6. De-centralized.** there is no need for trusted centralized point, Blockchain assures anyone can play the role of ledger.

**7. Distributed.** data can be stored on various data sources. Since our system separates actual data store from data management, no data has to be moved from its point of origin.

**8. Asynchronous.** both data and service are always available to clients with bad connectivity. Pub-Sub ensures messages are cached by the Publishers

Table 2: Overhead of create and verify messages in the data management protocol.

| Step | Role | Quantity | # Processing Cores | Processing Overhead (second) |
|---|---|---|---|---|
| Message 1 (Create) | Source | N/A | 1 | $1.20E-02$ |
| Message 1 (Verify) | Owner | N/A | 1 | $2.00E-02$ |
| Message 2 (Verify) | Owner | 1 Endorser | 1 | $1.66E-02$ |
| Message 2 (Verify) | Owner | 10 Endorsers | 1 | $2.20E-01$ |
| Message 3 (Create) | Owner | 1 Capability | 1 | $8.77E-03$ |
| Message 3 (Create) | Owner | 10 Capabilities | 1 | $8.57E-02$ |
| Message 3 (Create) | Owner | 100 Capabilities | 1 | $8.54E-01$ |
| Message 3 (Verify) | Requester | 1 Capability | 1 | $2.37E-02$ |
| Message 3 (Verify) | Requester | 10 Capabilities | 1 | $1.89E-01$ |
| Message 3 (Verify) | Requester | 100 Capabilities | 1 | $1.84E+00$ |
| Message 4 (Create) | Requester | 1 Capability | 1 | $1.59E-04$ |
| Message 4 (Create) | Requester | 10 Capabilities | 1 | $3.98E-04$ |
| Message 4 (Create) | Requester | 100 Capabilities | 1 | $2.79E-03$ |
| Message 4 (Verify) | Source | 1 Capability | 1 | $3.51E-02$ |
| Message 4 (Verify) | Source | 10 Capabilities | 1 | $3.28E-01$ |
| Message 4 (Verify) | Source | 100 Capabilities | 1 | $3.26E+00$ |
| Message 4 (Verify) | Source | 100 Capabilities | 16 | $2.00E-01$ |
| Message 4 (Verify) | Source | 1000 Capabilities | 16 | $2.05E+00$ |
| Message 5 (Create) | Source | N/A | 1 | $1.02E-02$ |
| Message 5 (Verify) | Owner | N/A | 1 | $1.54E-02$ |

and Blockchain assures the availability of data.

**10. Transparent.** this is a property of Blockchain.

**11. Auditable.** another property of Blockchain.

The following are general properties of our system:

**9. Compliant.** our model provides a particular advantage, because it does not store any data, or user identifiable information, it may not be subject to the regulatory requirements.

**13. Usable.** all the data owner functionalities are exposed to a user through a mobile app, then can simply view their data portfolio and grant access to data requesters.

## 8 Related Works

## Appendix I: Data Management Protocol Analysis

Security correctness is a major requirement for practical applications. In Dusc, this requirement is assured thorough using delegation by certificate method. In this section we provide a more formal proof using Abadi et al's notion and method introduced in [2] and [3].

The data management protocol relies on two abilities:

D1. Ability of the Source (S) to give to another principal Owner (O) the authority to act on Source's behalf

D2. Ability of the Owner (O) to give to another principal Requester (R) the authority to act on Owner's behalf

In this section, these abilities are discussed as a form of delegation (more throughly studied in [2]). In our delegation model, when a data requester ($R$) requests a data from a data source ($S$), $R$ presents a token demonstrating: $R$ is making the request and owner ($O$) has delegated to $R$. Upon $R$s request, an access will be granted, only if $S$ decides this delegation is permitted. This process is usually done by looking up an ACL. However, in our system $S$ delegates the decision on the former delegation to $O$ through another delegation as a substitution of an ACL record. In other word, $O$ is delegated to decide if delegation of $O$ to $R$ is permitted to access a certain object. This positions $O$ in the center of access control.

In the proposed protocol these rights in each delegation is as follows:

1. in *D1* Source delegates the access control decision on a certain data object to Owner

2. in *D2* Owner delegates the access to this data object on source with a certain query to the requester

In our system, delegation certificates are not transferable, and access control decisions depends only on the identities of both $O$ and $R$.

Delegation relies on some form of trust among $R$, $O$, and $S$ preferably by authentication. In our system:

1. Owner and source are assumed authenticated (Assumption A1, A2)

2. Owner authenticates endorsers by using a trusted third party and may trusts the requester by endorsement (Assumption A3)

We consider three instances of delegation. In each case we are led to ask whether composite principals, such as B—A, appear on Cs ACL. The simplest instance of delegation is delegation without certificates:

1. When B wishes to make a request r on As behalf, B sends the signed request along with As name, for example in the format KB says A says r

2. When C receives the request r, he has evidence that B has said that A requests r, but not that A has delegated to B; then C consults the ACL for request r and determines whether the request should be granted under these circumstances

The following is the reasoning of the data source (S), who makes the access control decision.

In the message 4, when $R$ wishes to access the data object ($d$) on the data owner ($O$)'s behalf, the data requester ($R$) sends the data access ticket ($DAT$) received in step 3, signed with its public key ($K_R$) to the data source ($S$). It contains two delegation certificates:

1. *DOT* which is the certificate of *D1* deligation.

2. $\{Query, O, K_O^3, K_R\}_{K_O'}$: which is the certificate of *D2* deligation.

The signed message of $\{Query, O, K_O^3, K_R\}_{K_O'}$ verifies that $R$ is deligated by $O$ under $R \wedge Query \wedge d$ access right:

$$O \text{ says } [R \wedge Query \wedge d])$$

At this point, $S$ needs to decide if this deligation have access to data object $d$. Using *DOT*, $S$ verifies that $O$ is deligated to control $d$ by $S$ itself:

$$S \text{ says } (O \text{ controls } d)$$

Therefore access will be granted.